# COMMENTS ON "A NEW TRANSIENT ATTACK ON THE KISH KEY DISTRIBUTION SYSTEM"

**Laszlo B. Kish** [1), ] **Claes G. Granqvist** [2)]

[1)] *Texas A&M University, Department of Electrical and Computer Engineering, College Station, TX 77843-3128, USA*

[2)] *Department of Engineering Sciences, The Ångström Laboratory, Uppsala University, P.O. Box 534, SE-75121 Uppsala, Sweden*

**Abstract**

A recent IEEE Access Paper by Gunn, Allison and Abbott (GAA) proposed a new transient attack against the Kirchhoff-law–Johnson-noise (KLJN) secure key exchange system. The attack is valid, but it is easy to build a defense for the KLJN system. Here we note that GAA's paper contains several invalid statements regarding security measures and the continuity of functions in classical physics. These deficiencies are clarified in our present paper, wherein we also emphasize that a new version of the KLJN system is immune against all existing attacks, including the one by GAA.

Keywords: Measurement theory; Information security; Foundations of physics; Engineering over-simplifications.

## 1. Introduction

Research and development on unconditionally secure communication and key exchange have a history of progress via attacks and debates and, for example, this type of evolution has taken place for quantum key distributions (QKDs) [1,2, and references therein]. The present paper concerns the classical statistical-physics-based Kirchhoff-law–Johnson-noise (KLJN) key distribution system, delineated in Figure 1, which is no exception to the tradition of the research area, and the creation of the KLJN schemes [3,4] immediately triggered attacks [5-7]. The various attacks [5–16] have led to useful discussions [17–23], including corrections of flaws in the attacks [19–23] and developments of new defense protocols [5,10,11,13,24,25] as well as protocols that have increased immunity against attacks in general [24–27]. Furthermore, KLJN schemes that are totally immune to a certain attack have been presented [13,28–30] as has a new system that is immune to *all* existing attacks [31]. Responses to the attacks have included plain denials of their validity [18,21–23], and in some cases experimental results that purportedly supported an attack have been found flawed [23]. The debates sometimes represent a standoff between opposing parties with different scientific backgrounds, which is a typical feature of science debates on breakthrough results in physics, as observed already by Max Planck [32].

Recently, Gunn, Allison and Abbott (GAA) published an interesting paper [15] with the first attack utilizing transients at the beginning of the bit-exchange. Their idea is impressively simple and involves monitoring the mean-square voltage before the front of the transient reaches the other end of the communication cable. We note that this approach requires a very short sampling time—less than 10% of the correlation time for the noise [14]—and the relative change of the voltage is typically small during this period.

In a simple illustration of the key effect of GAA's approach, we assume that Eve monitors the voltage on the cable while its capacitance $C$ is charged up by a DC voltage via a resistor $R$. According to the Johnson–Nyquist formula [3], the voltage noise spectrum can be written as $S(f) = 4kTR$—where $f$ is frequency, $k$ is Boltzmann's constant and $T$ is temperature—which means that the larger resistance has a higher mean-square voltage. Thus the DC voltage scales with $\sqrt{R}$, whereas the $RC$ time constant scales linearly with the resistance and the rate scales inversely with this time constant. If Alice and Bob use no precaution and abruptly switch the resistors (with their generators) to the line, then the mean absolute value of the rate-of-change for cable voltage at the entry point will scale as $\sim 1/\sqrt{R}$. It is also obvious from the above considerations that a linear ramping-up of the noise amplitude is not helpful, at least not if the communicating parties perform the ramping in a symmetrical fashion as in the first experimental demonstration [11] of the KLJN scheme.

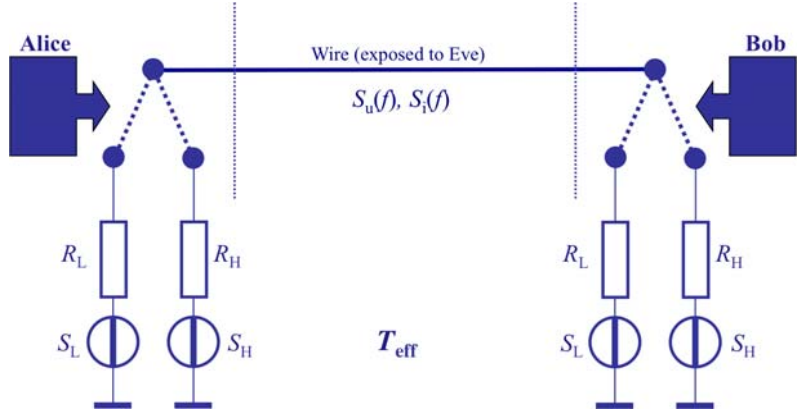

FIGURE 1. Outline of the KLJN scheme without defense circuitry [3] against active (invasive) attacks and attacks utilizing non-idealities. The $R_L$ and $R_H$ resistors, identical pairs at Alice and Bob, represent the Low (*L*) and High (*H*) bit-values. The corresponding (band-limited) white noise spectra $S_L$ and $S_H$ form identical pairs at the two ends, but they belong to independent Gaussian stochastic processes. Both parties are at the same temperature $T_{eff}$, so the net power flow is zero. The *LH* and *HL* bit-situations of Alice and Bob produce identical voltage and current noise spectra, $S_u$ and $S_i$, in the wire, implying that they represent a secure bit exchange. The total loop resistance $R_{loop}$ is publicly known and can be calculated by the measured voltage noise or current noise spectrum and the Johnson formula, for example as $R_{loop}=4kT/S_i$. In the *LH* and *HL* case, Alice and Bob can calculate the resistance at the other end of the cable by subtracting their own resistance value from $R_{loop}$. The *LL* and *HH* bit arrangements, which occur in 50% of the cases, do not offer security. Consequently 50% of the bits must be discarded. This system works also with arbitrary, continuum resistor values to securely generate and share continuum random numbers.

We have confirmed GAA's conclusion [15] that their attack works with about $p = 0.7$–$0.8$, where $p$ is Eve's probability of successfully guessing the key-bits. These values of $p$ require four stages of the simplest XOR-based privacy amplification in order to reduce $p$ to its ideal range of $0.5 < p < 0.5006$ [33], which implies a corresponding 16-fold slowdown of the key exchange. Then the corresponding relative information leak toward Eve is less than $10^{-8}$ [33].

## 2. A KLJN scheme that is immune against the attack

There are many easily realizable ways to reduce the efficiency of GAA's attack [15], and some of them were outlined in their paper. Here we emphasize that the new Random-Resistor–Random-Temperature (RRRT) KLJN scheme [31], see Figure 2, is *totally immune* against not only GAA's recent attack but against *all* presently known attacks.

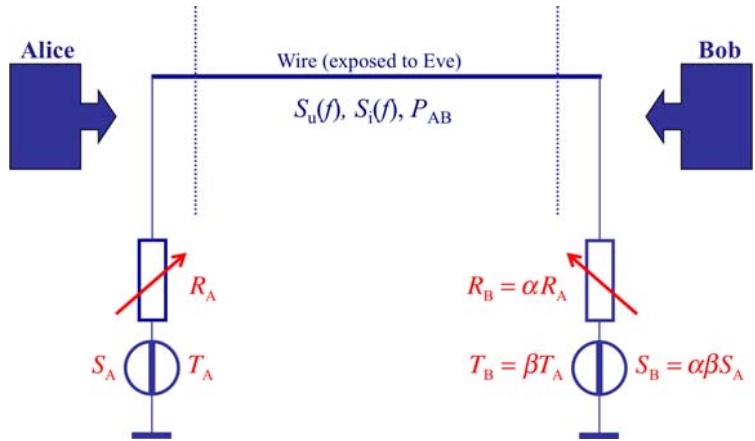

FIGURE 2. Outline of the RRRT–KLJN scheme [31]. The temperatures and resistors at Alice (A) and Bob (B), and their corresponding voltage noise spectra, are continuum random variables with a new random choice made at the beginning of each KLJN period. The Low (*L*) and High (*H*) bit values at Alice and Bob are determined by relative resistance values; for example, the party with the higher resistance has the high bit. Eve cannot determine even the sum of the resistance values, because there is no public knowledge about the temperatures, and the mean-square noise voltages and the resistances fluctuate independently. (Various symbols are defined in Figure 1 and elsewhere [31].)

The mean-square generator voltages and the resistance values are independent random variables, and hence Eve cannot relate the measured transient mean-square voltage to the resistance value. Thus GAA's transient attack [15] yields zero information leak about the key ($p = 0.5$).

## 3. Deficiencies in the GAA paper

The main reason for writing our present article is that, although GAA's attack works, there are important deficiencies in their paper [15], which we want to correct. A general comment is that GAA's paper is poorly documented and void of details regarding simulations: for example, what cable model or software was used, how were the noise generated, what simplifications were assumed, etc? All of these questions point at essential information needed for assessing the validity of, and and potential flaws in, GAA's simulations [14,22–24].

We confirmed GAA's results [15] by use of the LTSPICE industrial cable simulator (details of these types of simulations, and their underlying assumptions, are described elsewhere [14]), and in this section we focus on remaining problems inherent in GAA's work [15]. First, in subsections A and B, we address two minor mistakes, the second one possibly emanating from an inadequacy in our former paper [24], which we also correct here. Finally, in subsection C, we deal with major flaws in GAA's paper [15] regarding physics and security claims based on incomplete circuit theory.

### A.  Secrecy rate as a measure of unconditional security

GAA [15] use the term "secrecy rate" to characterize security. We note that GAA cite several old papers concerning secrecy rate [4,17,34,35], but this term does not exist in the mentioned papers. For a modern discussion of "secrecy rate", we refer to work by Chorti and Poor [36].

When using secrecy rate to characterize security, the bit-error probability $q$ of Alice and Bob enters into the result: the higher the $q$, the lower the secrecy rate. Under certain special conditions, but not for KLJN in general, one can relate the secrecy rate to the maximum rate of secure bit-exchange after privacy amplification. However security (secrecy) and secrecy rate are like apple and orange: both are fruits but are completely different. Secrecy rate is not an ultimate measure of security but is only a practically useful performance parameter in some special situations of secure communication [36]. The basic condition of (perfect unconditional) secrecy of key exchange is that the probability of Eve's successful guessing of the key-bit is not improved by her monitoring of the key exchange [37,38].

When Eve's knowledge of a uniformly generated key is zero, then her probability of successfully guessing the bits is 0.5. In the case of perfect secrecy (security), $p$ remains at this value even when Eve is eavesdropping, and the bit-error probability of key exchange between Alice and Bob is irrelevant. We offer the following illustrative example to show the inadequacy in using secrecy rate to judge the security of the KLJN system [39,40]: By manipulating wire resistance, frequency bandwidth and bit-detection thresholds, it is possible to design two different KLJN systems with identical secrecy rate; one of the systems has very poor security ($p \approx 1$) while the other has very strong security ($p \approx 0.5$).

One should note that, in the case of the KLJN system, the bit-error probability can be so low that it is not even measurable, such as $10^{-20}$, and therefore the use of secrecy rates may lead to the same conclusions as when $p$ is employed. However, it is incorrect to use secrecy rate in order to characterize the level of unconditional security of key exchange, and such an error can produce misleading results.

### B.  Parameter tuning to approach perfect security

The Appendix of GAA's work [15] contains some incorrect comments about our general security proof [24] for the KLJN key exchange. Here we discuss a minor issue: how to reach a desired security level by properly tuning the parameters of the KLJN system to be sufficiently close to their ideal values. For a mathematical proof of unconditional security, it is not necessary that this tuning is economical or practical—the tuning merely has to be physically achievable.

We first note that GAA cites the classical Diffie-Hellman paper [41] (their reference [2]) about unconditional security. This paper is from the times when physical unconditionally secure key exchange did not yet exist and the key was supposed to be perfectly unconditionally secure (such as by delivery via courier) or only conditionally secure by using one-way-functions. Unconditional security of physical key exchangers, on the other hand, was introduced [1] 25 years later by Mayers [42]. An unconditionally secure physical key-exchanger is never perfectly secure, but perfect security can be approached arbitrarily at least conceptually.

GAA argue [15] correctly that, when parameters are tuned toward their ideal values to match security requirements, there are some parameters that may have limits for doing that. GAA use the example of cable length, which can rarely be close to zero (except in intra-instrument chip-to-chip communication). This limitation is true, but the goal of a proof for unconditional security is to show that there is a set of practical parameter values for which the required security level is reached. This was proved by us [24] via the continuity of functions in stable classical-physical systems, which implies that the parameters approach their ideal value when *p* converges towards 0.5 representing perfect security. However, *all* parameters are not required to approach their ideal values for convergence to perfect security. For example, the influence of a large cable-length can be evaded by privacy amplification [33] at the cost of a sufficiently small bandwidth (the more privacy amplification steps, the smaller the bandwidth), and thus one can reach the required security level even when the cable-length is significant. Clearly, invested time is the ultimate price to pay in order to approach perfect security, which is the same as in QKD [1].

We now illustrate the case of a finite cable-length with the hypothetical function $p = 0.5 + xy$, where *x* represents the cable-length and *y* the reduced bandwidth (reciprocal of the time duration of bit-exchange). Perfect security is approached when $x \to 0$ and/or $y \to 0$. Consequently, it is not necessary that both *x* and *y* approach their ideal values; one parameter can stay significant and the system still converges to perfect security. Practical situations are of course more complex and less ideal than in this example.

Finally, we note that our previous paper [15] contains an inadequacy in Equation 5, where the Taylor polynomial is shown only up to first order. However most effects in the KLJN system require an expansion at least up second order and this is so for example in the example above. However, the inadequacy does not alter the main conclusion about the existence of unconditional security and the fact that the fundamental base of unconditional security of the KLJN system lies in the continuity of functions in stable classical-physical systems and in the existence of a parameter set, belonging to perfect security, which can be approached in a continuum fashion.

## C. Continuity of functions in stable classical-physical systems

The last and most important point of concern about GAA's paper [15] deals with security versus physics. As an objection to our earlier argument [24] that *p* can be tuned in a continuous fashion in KLJN, GAA claim in the Appendix of their article [15], by using a circuit example, that functions in stable classical-physical systems are not always continuous.

This is an assertion with very far-reaching consequences! However, it must be incorrect since otherwise the whole theoretical framework, as well as the education, of classical physics—including mechanics, elasticity, electrodynamics, fluid dynamics, statistical physics, condensed matter physics, etc. [43–50]—are flawed. We therefore investigate GAA's claim [15] and argue that there are three different types of errors in their argumentation; they are related to electrical circuit theory, security and physics.

We first note that the system examined by GAA [15] is *not* the KLJN system. The circuit underlying their demonstration is shown in Figure 3, which represents a situation wherein it is publicly known that Alice and Bob have 1-Ω resistors whereas their DC voltages $U_A$ and $U_B$ are secret. For the sake of simplicity, we assume that the arbitrary voltages $U_A$ and $U_B$ represent bit-values in a pre-agreed, realistic fashion. Eve measures the voltages $U_{AE}$ and $U_{BE}$. If $R_E > 0$, she can exactly determine the voltages $U_A$ and $U_B$ from the measured quantities, and thus Eve has perfect eavesdropping of the bit-values (*p* = 1). In the case of $R_E = 0$, see Figure 4, Eve cannot determine the secret voltages from $U_{AE}$ and $U_{BE}$ because the related matrix is not invertible, and then GAA [15] argue that Eve has zero information about the bits (*p* = 0.5). From this fact, GAA conclude that the transition from complete information with *p* = 1 (at $R_E > 0$) to zero information with *p* = 0.5 (at $R_E = 0$) proves the existence of a non-continuous $p(R_E)$ function because of the singularity of *p* at $R_E = 0$. Thus GAA profess that a function related to security is non-continuous in a classical-physical system.

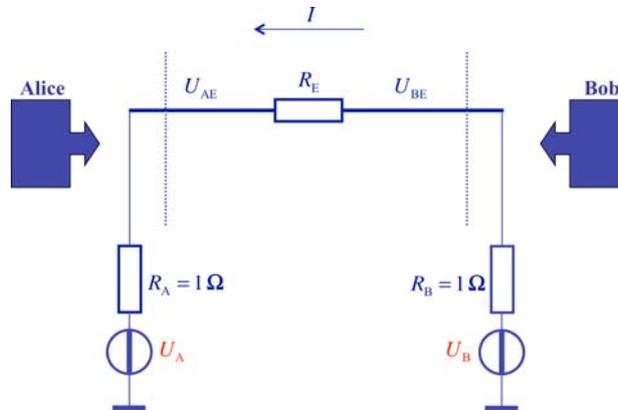

FIGURE 3. Outline of the situation wherein Eve knows the resistor values and measures the voltages $U_{AE}$ and $U_{BE}$. If $R_E > 0$, she can determine the voltages $U_A$ and $U_B$.

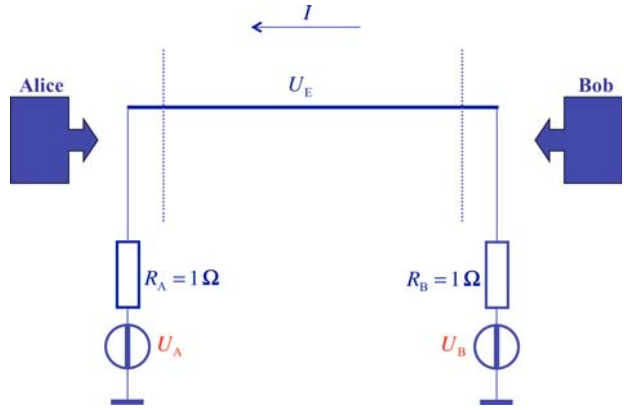

FIGURE 4. Outline of the situation wherein Eve measures voltage and knows the resistor values but still cannot find out the voltages. However, this is an absurdly limited Eve because she does not measure current, which contradicts the rules of security that Eve must utilize all of the information she has access to. If she measures also current, she can determine the voltages.

The following questions emerge as a result of GAA's claims:
(*i*) When GAA [15] discuss security and eavesdropping, do they follow the elementary rules of attacks against unconditional security, in particular, do they exploit all of Eve's available circuit measurement tools in the attack?
(*ii*) Is GAA's approach physical or is it only an unphysical engineering-type simplification from which one cannot draw safe conclusions about underlying physics?
(*iii*) If GAA's approach is indeed unphysical, does their conclusion hold if their approach is modified so as to make it physical?

## 4. Discussion

We now scrutinize the questions (*i*)–(*iii*). The answer to question (*i*) is that by ignoring the possibility of current measurement one makes a circuit-theoretical mistake that leads to an absurdly limited Eve. This contradicts the basic rules of security analysis that Eve must utilize all of the information she has access to. If she measures also current $I$, as can be done in various ways, she can determine the voltages exactly by $U_A = U_E - I*1\Omega$ and $U_B = U_E + I*1\Omega$. Clearly the role a non-zero $R_E$ is to enable current to be used in order to provide extra information via the voltage drop over $R_E$ and, equally obviously, this information is lost at $R_E = 0$. Consequently, there is no discontinuity within GAA's approach [15]. In fact, Eve's eavesdropping ability is constant and maximum ($p = 1$), and hence the situation explored by GAA does not offer security.

The reply to question (*ii*) is that the system GAA investigate [15] in order to challenge a fundamental rule of classical physics is unphysical because their circuit model does not contain the Johnson noise voltage sources of the resistances. This deficiency implies an underlying assumption about the physical system, *viz.*, is its being at zero absolute temperature. However, zero absolute temperature cannot be reached as a consequence of the Laws of thermodynamics and statistical physics, and assuming its existence renders GAA's model unphysical.

To answer question (*iii*), finally, we make GAA's system physical by adding Johnson noise generators $U_{An}(t)$, $U_{Bn}(t)$ and $U_{En}(t)$ to the corresponding resistors; see Figure 5. The impact of the non-zero noise is pervasive, and the Gaussian distribution of Johnson noise voltage guarantees non-perfect eavesdropping and a continuous transition of $p$ toward $R_E = 0$. The amplitude density of a Gaussian process will never reach zero and thus, mathematically, noise can cause bit-flips at any finite value of $U_A$ and $U_B$. For $R_E \to 0$, the DC voltage drop and noise on $R_E$ scale with $R_E$ and $\sqrt{R_E}$, respectively, which implies that the relative inaccuracy of the measured voltage on $R_E$ scales with $1/\sqrt{R_E}$ and is divergent when approaching the limit of $R_E = 0$; this divergence takes place in a continuum fashion. Perfect eavesdropping at non-zero $R_E$ can be achieved only via infinitely long time-averaging, which is an unphysical situation. Consequently, $p$ is a continuous function of $R_E$ at finite-time averaging and non-zero temperature,

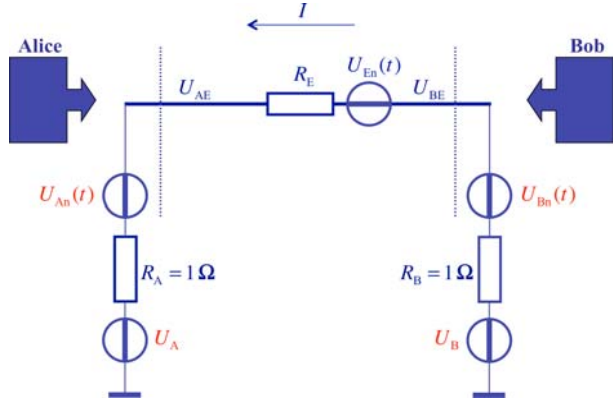

FIGURE 5. Outline of the situation encompassing correct physics represented by non-zero Johnson noise voltages of the resistors. The impacts of $U_{An}$ and $U_{Bn}$ are significant: for GAA's absurdly limited Eve it provides non-perfect eavesdropping and a continuous change of the information leak when $R_E$ converges to zero. Perfect eavesdropping can be achieved only after infinitely long time-averaging, which is unphysical.

## Conclusions

We have analysed a recent paper entitled "A New Transient Attack on the Kish Distribution System" by Gunn, Allison and Abbott [15]. Their attack is valid, but countermeasures are readily found. Our present paper discusses the arguments behind the attack, and we found a number of flaws that are sufficiently general to warrant a detailed treatment, as presented above. In particular, GAA's "proof" of the existence of a discontinuous function in a stable, linear, classical-physical system is invalid. Continuum functions prevail as soon as the system is made physical by including unavoidable thermal fluctuations. Our analysis shows, once again, that *over-simplified engineering models are unable to prove or disprove the Laws of Physics*.

## Appendix

This manuscript, in its various versions, has been criticized by reviewers and others on issues of a fundamental character. Some of these issues, we believe, are connected with traditions and lines of thought entrenched in different disciplines, such as in Engineering and in Physics. Here we give a brief discussion of some of the contentious items that may be of general relevance and where, as we perceive the situation, the engineering tradition leads to over-simplified, or even erroneous, results. We underscore that the present exposition is preliminary and awaiting a more in-depth treatment.

Our presentation in this paper rests the existence of *continuous functions*. It has been questioned that such functions are important for the present purpose, and it has been argued that we are guilty of having invented an "imaginary law" without giving any reference. However, it is an elementary issue that, in classical physics, functions describing linear systems and stable non-linear systems must be continuous. This is the very reason why differential calculus can be applied throughout classical physics [43–50]! The reliance on continuous functions is usually not stated explicitly, but it is as basic as the assumption that calculations in physics follow the rules of algebra.

As a history aside, we note that the realization that continuous functions describe physics is underlying Newton's formulation of differential and integral calculus. Discontinuities simply do not exist in classical physics, and if a model says so it is over-simplified.

We illustrate our view with an example: High-school physics says that, when we heat ice, after reaching the melting point the temperature remains constant during continuous energy influx until the energy representing the latent heat of the ice is exceeded. At first sight this represents a discontinuity. But of course this is not so and statistical thermodynamics tells that the temperature and the latent heat are continuous functions associated with thermal fluctuations at the phase change. In fact, the same effect constitutes a vigorous research field in superconductivity, *viz.*, fluctuation-conductance. Other examples could be given.

An asserted counter example to the continuity of functions in physics, which was put forward by a reviewer, is related to the *special case of Euler's disc*, which is a well-known system in classical physics. A practical example of Euler's disc is a coin spinning on a flat surface. This object oscillates with increasing frequency and then suddenly stops, seemingly via a discontinuous process.

But what does "suddenly stops" mean here? According to Newton's Second Law, the abrupt cessation of non-zero motion of non-zero mass requires an infinitely strong and infinitely narrow pulse of force—*i.e.*, a Dirac pulse—which is unphysical although commonly employed in simplified calculations in electrical engineering in order to estimate the behaviour of circuits. However such simplified engineering-type estimations are insufficient to address fundamental questions in physics.

A deeper investigation of the physics related to Euler's disc shows that it never really stops. Its centre-of-mass and all of its molecules will continue to oscillate randomly to satisfy Boltzmann's Equipartition Theorem, which states that there is $kT/2$ mean energy per thermal degree of motion, where $k$ is Boltzmann's constant and $T$ is absolute temperature. Thus thermal noise guarantees that continuity prevails even when a simplified model may predict a discontinuity.

Another bone of contention regards the fact that GAA's paper, discussed by us above, purportedly highlights *discontinuities in a probability function*. As an example, it was argued that a stochastic physical system subjected to some limiter or threshold would have a truncated distribution (*i.e.*, a discontinuity), and therefore GAA's paper would be perfectly valid.

But this assertion is flawed. Probability functions are always continuous in any physical system, and this includes not only classical physics but also quantum physics. Freshman quantum physics of quantum tunnelling serves as a good example: This treatment uses the fact that wave functions and their squared absolute values, which are the probability density of the particle under consideration, are always continuous whenever the height of the potential barrier is finite, *i.e.*, for any physical system. The underpinning reason for this can be found in Schrödinger's Equation. There are certainly discontinuous energy solutions in solid-state quantum systems, but they represent different states and continuity persists within any single state. The only discontinuities of probability in quantum physics happen during quantum measurements.

Finally, a reviewer claimed that "a stochastic physical system that is subjected to some limiter or threshold will have a truncated distribution". But this is obviously unphysical, and no physical limiter or threshold can be mathematically abrupt. There is always a continuous transition at the level of the limitation, which is evident since an abrupt limit would require not only infinite power but also infinitely fast response—and both requirements are unphysical. The analogy to Euler's disc is evident.

Our main conclusion, as we have stated also in the past, is that *over-simplified engineering models are unable to prove or disprove the Laws of Physics*.

# Acknowledgements

Valuable discussions with Horace Yuen, Vincent Poor, Tamas Erdelyi and Laszlo Leindler are appreciated.